\documentclass[reprint,amsmath,amssymb,aps,pra]{revtex4-2}
\usepackage[colorlinks,linkcolor=blue,anchorcolor=blue,citecolor=blue,urlcolor=blue]{hyperref}
\usepackage{txfonts}
\usepackage{graphicx,hyperref}
\usepackage{subfigure}
\usepackage{dcolumn}
\usepackage{bm}
\usepackage{braket}
\usepackage{upgreek}
\usepackage{extarrows}
\usepackage{setspace}
\usepackage{float}
\usepackage{rotating}
\allowdisplaybreaks[4]
\begin{document}
	\title{Quantum-classical correspondence of non-Hermitian spin-orbit coupled bosonic junction}
	
	\title{Quantum-classical correspondence of non-Hermitian spin-orbit coupled bosonic junction}
	\author{Xin Yan$^{1}$}
	\author{Hongzheng Wu$^{1}$}
	\author{Changwei Fan$^{1}$}
	\author{Baiyuan Yang$^{2}$}
	\author{Yu Guo$^{3}$}
	\author{Xiaobing Luo$^{1}$}
	\altaffiliation{Corresponding author: xiaobingluo2013@aliyun.com}
	\author{Jinpeng Xiao$^{2}$}
	\altaffiliation{Corresponding author:  xiaojinpeng2018@163.com}	
	\author{Zhao-Yun Zeng$^{2}$}
	\altaffiliation{Corresponding author:  zengzhaoyun@jgsu.edu.cn}	
	\affiliation{$^{1}$Department of Physics, Zhejiang Sci-Tech University, Hangzhou, 310018, China}
	\affiliation{$^{2}$School of Mathematics and Physics, Jinggangshan University, Ji'an, 343009, China}
	\affiliation{$^{3}$Hunan Provincial Key Laboratory of Flexible Electronic Materials Genome Engineering, School of Physics and Electronic Science,
		Changsha University of Science and Technology, Changsha 410114, China}        
	\date{\today}
	\begin{abstract}
		We investigate the classical-quantum correspondence of non-Hermitian Spin-orbit (SO)-coupled bosonic junctions, where an effective decay term is introduced in one of the two wells. Starting from the normalized two-point functions, we analytically demonstrate that the mean-field system has a classical Hamiltonian structure, and we successfully derive a non-Hermitian discrete nonlinear Schrödinger (Gross-Pitaevskii) equation. We discover that near the symmetry-breaking phase transition point, the correspondence between classical (mean-field) and quantum dynamics is more likely to break down. When the effective spin-orbit coupling (SOC) strength assumes half-integer values, atomic self-trapping in the non-lossy well definitely occurs, regardless of the system parameters, and the quantum dynamics is insensitive to the number of particles. Additionally, we reveal that in both the mean-field and many-particle models, the SOC effects can greatly promote the synchronous periodic oscillations between the spin-up and spin-down components, and this synchronization dynamics is protected by a symmetry mechanism.
	\end{abstract}
	\maketitle
	\section{Introduction}
	Spin-orbit coupling is a fundamental quantum phenomenon that couples a particle’s intrinsic spin with its orbital angular momentum, and it finds applications in various fields, including spintronic devices, topological insulators, quantum computing, and high-precision measurements \cite{3,4}. Experimentally, SOC has been successfully implemented in Bose-Einstein condensates (BECs) through artificial gauge fields \cite{8,9,10,11,12}. Given that the double-well potential is a standard model for studying tunneling and Josephson physics, SO-coupled BECs, where two internal atomic states are coupled to each other via a pair of counterpropagating laser beams, have received considerable attention when placed in such potentials. To date, research on this topic has been conducted from the viewpoints of both full-quantum and mean-field treatments, with particular emphasis on the dynamics of Josephson oscillations\cite{33,67,74,75,76,79,80}, self-trapping\cite{78}, flat bands in the energy spectrum\cite{66}, and dynamical suppression\cite{83,84,85,86,101}.
	Here, two hyperfine atomic states are treated as spin (pseudospin), which serve as additional internal degrees of freedom, thereby opening up new possibilities for investigating the Josephson effects.
	
	In recent years, the study of open systems has garnered significant interest among researchers. Experimentally, open Bose-Einstein condensate (BEC) systems have been effectively created, and the dissipation can be precisely engineered \cite{57}. The theoretical frameworks for open systems encompass the Lindblad master equation and non-Hermitian Hamiltonians. Non-Hermitian Hamiltonians are introduced as a phenomenological approximation theory, and their implementation is achievable using the Feshbach projection operator method \cite{91}. A rich variety of novel phenomena can emerge in open systems \cite{16,19,20}. Among them, non-Hermitian Hamiltonians with parity-time ($\mathcal{PT}$) symmetry are particularly intriguing, as they exhibit a symmetry-breaking phase transition at which the spectrum changes from all real (the $\mathcal{PT}$ phase) to
	complex (the $\mathcal{PT}$-broken phase) when the non-Hermitian parameter exceeds a certain threshold \cite{22}. Recently, the connection of the non-Hermitian physics to SO-coupled atomic systems has been explored, focusing primarily on ground states \cite{99,100}, steady-state analytical solutions \cite{25}, phase transitions, and exceptional points (EPs) \cite{16,17,18,90}. 
	The Floquet control of $\mathcal{PT}$-symmetric SO-coupled noninteracting BECs in a double-well potential has also been investigated \cite{70}. A more encouraging advancement is that the role of dissipation in SO-coupled atomic systems has been studied experimentally \cite{97}, which demonstrates that non-Hermitian topological states can be controlled with SOC.
	
	The quantum-classical correspondence (QCC) bridges the gap between macroscopic and microscopic perspectives, carrying significant theoretical and practical value. For Hermitian systems, the correspondence between the many-particle description and the mean-field approximation can be investigated in analogy with the usual quantum-classical correspondence for single-particle quantum mechanics. In particular, the mean-field dynamics of the Bose-Hubbard model and the QCC in this context have been extensively studied \cite{33,34,35,64,71,98}.
	Later, considerable attention has been paid to the open systems, and a series of mean-field results and beyond-mean-field approximations of the dissipative Bose-Hubbard model, based on a master equation description, have been reported in the literature \cite{43,44}. In parallel, there are numerous studies on the QCC of non-Hermitian Bose–Hubbard dimers \cite{13,14,15,40,41,63}, where a complex Hamiltonian models a BEC in a double-well potential experiencing an effective decay from one of the modes. Such a purely decaying Hamiltonian can be mapped to a $\mathcal{PT}$-symmetric version including a source and sink of equal strength, and the corresponding classical (mean-field) dynamics can then be derived by investigating the quantum evolution of the SU(2) expectation values of the three angular momentum operators in the large $N$ limit \cite{13,14,15,40,41}. However, due to the complexity of the non-Hermitian Bose-Hubbard model with SOC, we note that the non-Hermitian generalization of the discrete nonlinear Schrödinger equation resulting from the mean-field approximation has not been obtained, and the QCC in such systems has yet to be explored.
	
	In this paper, we investigate the quantum-classical correspondence (QCC) of a non-Hermitian SO-coupled BEC in a double-well potential (a non-Hermitian SO-coupled bosonic junction), where non-Hermiticity arises from leakage in the right well \cite{72}. Following the mean-field approximation methods from Refs.~\cite{43} and \cite{40}, we derive the equations of motion for the normalized two-point functions (also referred to as the single-particle reduced density matrix). We have conducted a rigorous mathematical proof confirming that the derived mean-field system embodies a classical Hamiltonian structure, and we have successfully derived the non-Hermitian version of the discrete nonlinear Schr\"odinger  (Gross-Pitaevskii) equation. We demonstrate the effectiveness of the aforementioned calculation method for generalized classical Hamiltonian functions in a wider class of extended non-Hermitian, $N$-particle, $M$-mode Bose-Hubbard systems. By mapping the purely decaying model to a $\mathcal{PT}$-symmetric version, we find that when the effective SOC strength assumes half-integer values, $\mathcal{PT}$-symmetry-breaking appears even for arbitrarily small non-Hermitian parameters on both the mean-field and full-quantum sides. We subsequently examine the correspondence between full quantum dynamics and mean-field dynamics across different parameter regions. It is discovered that near the symmetry-breaking phase transition point, the quantum many-body fluctuation effects become significant, and the correspondence between the mean-field and quantum dynamics fails. While the system always resides in the symmetry-breaking regime when the SOC strength assumes half-integer values, the quantum dynamics are insensitive to changes in the number of particles. Additionally, we find that in both the mean-field and full-quantum systems, an extremely small SOC strength can induce synchronous periodic oscillations of the spin-up and spin-down components.
	\section{MODEL EQUATION}
	\label{section 2}
	We consider a non-Hermitian many-particle spin-orbit coupled bosonic junction described by the Hamiltonian\cite{40,84,101}
	\begin{equation}\label{1}
		\begin{split}
			\hat{H}&=-J(e^{-i\pi\gamma}\hat{a}_{L\uparrow}^\dagger\hat{a}_{R\uparrow}+e^{i\pi\gamma}\hat{a}_{L\downarrow}^\dagger\hat{a}_{R\downarrow}+h.c.)+\Omega(\hat{a}_{L\uparrow}^\dagger\hat{a}_{L\downarrow}\\&+\hat{a}_{R\uparrow}^\dagger\hat{a}_{R\downarrow}+h.c.)+\sum_{j,\sigma}(\frac{g}{2N}\hat{a}_{j\sigma}^\dagger\hat{a}_{j\sigma}^\dagger\hat{a}_{j\sigma}\hat{a}_{j\sigma}\\&+\frac{g}{N}\hat{a}_{j\uparrow}^\dagger\hat{a}_{j\downarrow}^\dagger\hat{a}_{j\uparrow}\hat{a}_{j\downarrow})-i\beta\sum_{\sigma}\hat{a}_{R\sigma}^\dagger\hat{a}_{R\sigma},
		\end{split}
	\end{equation}
	where $\hat{a}_{j,\sigma}^{\dagger}$ ($\hat{a}_{j,\sigma}$) is the creation (annihilation) operator for the ground-state wave function with spin up ($\sigma = \uparrow$) or down ($\sigma = \downarrow$) in the left ($j = L$) or right ($j = R$) well, respectively, obeying the bosonic commutation relation $[\hat{a}_{i,\sigma},\hat{a}_{j,\sigma'}^{\dagger}] = \delta_{i,j}\delta_{\sigma,\sigma'}$. $\Omega$ represents the Raman coupling strength, $\gamma$ is the SOC strength, $J$ is the tunneling amplitude between the two wells, and $g$ is the strength of the on-site interaction. Note that the Hamiltonian (\ref{1}) commutes with the total number operator $\hat{N} = \sum_{j,\sigma} \hat{a}_{j,\sigma}^{\dagger} \hat{a}_{j,\sigma}$, and the additional imaginary part of the mode energy, $\beta$, describes the decay over time of the probability to find the entire many-particle ensemble in the four modes\cite{40}.
	
	Let us now introduce the mean-field description for the non-Hermitian Bose-Hubbard dimer, referred to as model (\ref{1}), in the spirit of the classical approximation as we take the limit of infinite $N$. For mathematical conciseness, we use $\hat{a}_1, \hat{a}_2, \hat{a}_3, \hat{a}_4$ to denote $\hat{a}_{L\uparrow}, \hat{a}_{L\downarrow}, \hat{a}_{R\uparrow}, \hat{a}_{R\downarrow}$ in model (\ref{1}), which are the annihilation operators for the four modes. Our mean-field treatment can be applied to the extended non-Hermitian Bose-Hubbard model with a number of particles $N$ and an arbitrary number of $M$ modes, including arbitrary inter-mode particle interactions. To begin with, we can write the Hamiltonian $\hat{H}$ of the extended non-Hermitian many-particle Bose-Hubbard model as follows:
	\begin{equation}\label{2}
		\begin{split}\begin{aligned}
				\hat{H}=&\hat{H}_h+\hat{H}_a+\hat{H}_{n},\\
				\hat{H}_h=&\sum_{i,j=1}^{M}h^h_{i,j}\hat{a}_{i}^{\dagger}\hat{a}_{j},\\
				\hat{H}_a=&\sum_{i,j=1}^{M}h^a_{i,j}\hat{a}_{i}^{\dagger}\hat{a}_{j},\\
				\hat{H}_{n}=&\frac{1}{N}\sum_{i,j=1}^{M}h_{i,j}\hat{a}_{i}^{\dagger}\hat{a}_{j}^{\dagger}\hat{a}_{i}\hat{a}_{j},
		\end{aligned}\end{split}
	\end{equation}
	where $\hat{H}_h$ represents the Hermitian single-particle term, $\hat{H}_a$ is the anti-Hermitian single-particle term, and $\hat{H}_n$ is the Hermitian interacting term. These parts combine to form the complete Hamiltonian of a many-body system and satisfy the following relation:  
	\begin{equation}\label{3}
		\begin{split}\begin{aligned}
				\hat{H}_h^{\dagger}=\hat{H}_h, \hat{H}_a^{\dagger}=-\hat{H}_a, \hat{H}_n^{\dagger}=\hat{H}_n.
		\end{aligned}\end{split}
	\end{equation}
	\quad It is important to note here that any non-Hermitian operator can be decomposed into the sum of a Hermitian part and an anti-Hermitian part, thus we do not impose additional requirements on the single-body terms. Using Eq. (\ref{3}), we derive that $h^h$, $h^a$, and $h$ possess the following properties:
	\begin{equation}\label{4}
		\begin{split}\begin{aligned}
				(h^h)^{\dagger}=h^h, (h^a)^{\dagger}=-h^a, (h)^{*}=(h)^{\mathrm{T}}=h.
		\end{aligned}\end{split}
	\end{equation}
	We readily observe that the model (\ref{1}) we are considering is a specific case of model (\ref{2}), with $M = 4$.
	
	For a general quantum system with the Hamiltonian operator \(\hat{H}\), which is not necessarily Hermitian, the evolution of a pure state \(|\Psi\rangle\) is determined by the Schr\"odinger equation: $id{|\Psi\rangle}/dt = \hat{H}|\Psi\rangle$. From this, we can derive the non-Hermitian generalized equation of motion for the expectation value of an operator $\langle\hat{A}\rangle:={\langle\Psi|\hat{A}|\Psi\rangle}/{\langle\Psi|\Psi\rangle}$ as follows,
	\begin{equation}\label{5}
		i\frac{d\langle\hat A\rangle}{dt}=\langle[\hat A,\hat H]\rangle+2\Delta(\hat A),
	\end{equation}
	where $\Delta(\hat A)=\langle\hat{H}_a\hat A\rangle-\langle\hat{H}_a\rangle\langle\hat A\rangle$. To derive the mean-field approximation, we start with the two-point functions $\sigma_{ij} := \langle\hat{a}_i^\dagger\hat{a}_j\rangle/N$, which are also known as the reduced single-particle density matrix. In the limit of large $N$, we assume that all particles occupy the same state and can be described by a single macroscopic wave function. The fully condensed states can be expressed as SU(M) coherent states, namely, $|\vec{x},N\rangle_c = \frac{1}{\sqrt{N!}} \left(\sum_{i=1}^{M} x_i \hat{a}_i^\dagger\right)^N |0\rangle,$ where $\vec{x} := (x_1, x_2, \ldots, x_M) \in \mathbb{C}^M$ and we refer to $x_i$ as the coherent state parameters, which can be physically interpreted as the amplitude of a single particle occupying the $i$-th mode.  The mean-field equations of motion can be obtained
	by replacing the expectation values of the relevant operators in the two-point functions with their values in SU(M) coherent states, namely,  $\sigma_{ij}:={\langle\hat{a}_i^\dagger\hat{a}_j\rangle_c}/N=\frac{_c\langle \vec{x},N|\hat{a}_i^\dagger\hat{a}_j|\vec{x},N\rangle_c}{N _c\langle \vec{x},N|\vec{x},N\rangle_c}={x_i^*x_j}/n$, where $n=|\vec{x}|^2$. In the following, unless otherwise specified, for the sake of simplicity in notation, we omit the subscript ``$c$" and use ``$\langle\rangle$" to refer to ``$\langle\rangle_{c}$", which denotes the expectation values in SU(M) coherent states. Accoding to Eq. (\ref{5}), the equations of motion for the two-point functions can be formulated by
	\begin{align}\label{6}
		i\frac{d\langle\hat{a}_i^{\dagger}\hat{a}_j\rangle/N}{dt}=&\frac{1}{N}\Big(\langle[ \hat{a}_i^{\dagger}\hat{a}_j,\hat H]\rangle+\Delta(\hat{a}_i^{\dagger}\hat{a}_j)\Big)\nonumber\\
		=&\frac{1}{N}\Big(\langle\hat{a}_{i}^{\dagger}[\hat{a}_{j},\hat{H}_h]\rangle+\langle\hat{a}_{i}^{\dagger}[\hat{a}_{j},\hat{H}_a]\rangle+\langle\hat{a}_{i}^{\dagger}[\hat{a}_{j},\hat{H}_n]\rangle\nonumber\\
		+&\langle[\hat{a}_{i}^{\dagger},\hat{H}_h]\hat{a}_{j}\rangle+\langle[\hat{a}_{i}^{\dagger},\hat{H}_a]\hat{a}_{j}\rangle+\langle[\hat{a}_{i}^{\dagger},\hat{H}_n]\hat{a}_{j}\rangle\nonumber\\
		+&2\langle\hat{H}_a\hat{a}_i^{\dagger}\hat{a}_j\rangle-2\langle\hat{H}_a\rangle\langle\hat{a}_i^{\dagger}\hat{a}_j\rangle\Big).
	\end{align}
	Substituting Eq. (\ref{2}) into Eq. (\ref{6}), the resulting terms related to $\langle[\hat{a}_i^\dagger \hat{a}_j, \hat{H}]\rangle$ in Eq. (\ref{6}) are calculated as follows:
	\begin{align}\label{7}
		&\frac{1}{N}\langle\hat{a}_{i}^{\dagger}[\hat{a}_{j},\hat{H}_{h}]\rangle=\frac{1}{N}\sum_{l=1}^{M}h^{h}_{j,l}\langle a_{i}^{\dagger}\hat{a}_{l}\rangle=\sum_{l=1}^{M}h^{h}_{j,l}\frac{x_{i}^{*}x_{l}}{n},\nonumber\\
		&\frac{1}{N}\langle\hat{a}_{i}^{\dagger}[\hat{a}_{j},\hat{H}_{a}]\rangle=\frac{1}{N}\sum_{l=1}^{M}h^{a}_{j,l}\langle a_{i}^{\dagger}\hat{a}_{l}\rangle=\sum_{l=1}^{M}h^{a}_{j,l}\frac{x_{i}^{*}x_{l}}{n},\nonumber\\
		&\frac{1}{N}\langle[\hat{a}_{i}^{\dagger},\hat{H}_{h}]\hat{a}_{j}\rangle=-\frac{1}{N}\sum_{l=1}^{M}h^{h}_{l,i}\langle a_{l}^{\dagger}\hat{a}_{j}\rangle=-\sum_{l=1}^{M}h^{h}_{l,i}\frac{x_{l}^{*}x_{j}}{n},\nonumber\\
		&\frac{1}{N}\langle[\hat{a}_{i}^{\dagger},\hat{H}_{a}]\hat{a}_{j}\rangle=-\frac{1}{N}\sum_{l=1}^{M}h^{a}_{l,i}\langle a_{l}^{\dagger}\hat{a}_{j}\rangle=-\sum_{l=1}^{M}h^{a}_{l,i}\frac{x_{l}^{*}x_{j}}{n},\nonumber\\
		&\frac{1}{N}\langle\hat{a}_{i}^{\dagger}[\hat{a}_{j},\hat{H}_n]\rangle= 2\frac{N-1}{N}\sum_{l=1}^{M}h_{l,j}\frac{|x_{l}|^2}{n}\frac{x_i^*x_j}{n},\nonumber\\
		&\frac{1}{N}\langle[\hat{a}_{i}^{\dagger},\hat{H}_n]\hat{a}_{j}\rangle=- 2\frac{N-1}{N}\sum_{l=1}^{M}h_{i,l}\frac{|x_{l}|^2}{n}\frac{x_i^*x_j}{n},
	\end{align}
	and the results of the last two terms corresponding to $\Delta(\hat{a}_i^\dagger \hat{a}_j)$ in Eq. (\ref{6}) are given by
	\begin{align}\label{8}
		&\frac{1}{N}\big(2\langle\hat{H}^a\hat{a}_i^{\dagger}\hat{a}_j\rangle-2\langle\hat{H}^a\rangle\langle\hat{a}_i^{\dagger}\hat{a}_j\rangle\big)\nonumber\\
		&=2\sum_{k=1}^Mh^{a}_{k,i}\frac{x^*_kx_j}{n}-2\sum_{k,l=1}^{M}h^{a}_{k,l}\frac{x_{k}^{*}x_{i}^{*}x_{l}x_{j}}{n^2}.
	\end{align}
	In the derivation of Eq.~(\ref{8}), we have utilized the following relations: $\langle\hat{a}^{\dagger}_k\hat{a}^{\dagger}_i\hat{a}_l\hat{a}_j\rangle = N(N-1)\frac{x_{k}^{*}x_{i}^{*}x_{l}x_{j}}{n^2}$, $\langle\hat{a}^{\dagger}_k\hat{a}_l\rangle\langle\hat{a}^{\dagger}_i\hat{a}_j\rangle = N^2\frac{x_{k}^{*}x_{i}^{*}x_{l}x_{j}}{n^2}$, and $\langle\hat{a}_k^{\dagger}\hat{a}_l\hat{a}_i^{\dagger}\hat{a}_j\rangle = \delta_{i,l}\langle\hat{a}^{\dagger}_k\hat{a}_j\rangle + \langle\hat{a}^{\dagger}_k\hat{a}_i^{\dagger}\hat{a}_l\hat{a}_j\rangle$, where $\delta_{ij}$ is the Kronecker delta function. 
	
	To perform the mean-field approximation, we take the macroscopic limit $N \rightarrow \infty$ with $g$ fixed, and the last two terms of Eq. (\ref{7}) are given by
	\begin{equation}
		\begin{split}\begin{aligned}
				&\lim_{N\to\infty}\frac{1}{N}\langle\hat{a}_{i}^{\dagger}[\hat{a}_{j},\hat{H}_n]\rangle=2\sum_{l=1}^{M}h_{l,j}\frac{|x_{l}|^2}{n}\frac{x_i^*x_j}{n},\\
				&\lim_{N\to\infty}\frac{1}{N}\langle[\hat{a}_{i}^{\dagger},\hat{H}_n]\hat{a}_{j}\rangle=-2\sum_{l=1}^{M}h_{i,l}\frac{|x_{l}|^2}{n}\frac{x_i^*x_j}{n}.
		\end{aligned}\end{split}
	\end{equation}				
	
	Combining the above derivations, we obtain the classical coherent evolution of the two-point function:
	\begin{equation}\label{10}
		\begin{split}\begin{aligned}
				i\frac{d\sigma_{ij}}{dt}=&\sum_{l=1}^{M}(h^{h}_{j,l}+h^{a}_{j,l})\sigma_{il}-\sum_{l=1}^{M}(h^{h}_{l,i}+h^{a}_{l,i}-2h^{a}_{l,i})\sigma_{lj}\\
				-&2\sum_{k,l=1}^{M}h^{a}_{k,l}\sigma_{kl}\sigma_{ij}+2\sum_{l=1}^{M}(h_{l,j}-h_{i,l})\sigma_{ll}\sigma_{ij}.
		\end{aligned}\end{split}
	\end{equation}	
	Based on the equation (\ref{10}), we can already describe the classical (mean-field) behavior of the non-Hermitian many-particle system. 
	
	As an alternative to performing the coherent state approximation in the generalized equations of motion for the two-point function, we find that the mean-field description can also be expressed in terms of the generalized canonical equations of motion. Mathematically, in terms of the coherent state parameters $x_i$, the non-Hermitian generalization of the discrete nonlinear Schr\"odinger equation, which is equivalent to the evolution equation (\ref{10}), can be formulated as:                
	\begin{equation} \label{11}
		\begin{split}\begin{aligned}
				i\frac{dx_i}{dt}=\frac{\partial H}{\partial x_{i}^{*}}=\sum_{j=1}^{M}\widetilde{H}_{i,j}x_j,
		\end{aligned}\end{split}
	\end{equation}	
	where the classical Hamiltonian function is given by
	\begin{equation}\label{12}
		\begin{split}\begin{aligned}
				H=\lim_{N\to\infty}\frac{n\langle\hat{H}\rangle}{N}.
		\end{aligned}\end{split}
	\end{equation}
	In the canonical formulation of Eq. (\ref{11}), $x_i$ serves as the generalized coordinates, with $x_i^*$ acting as the conjugate generalized momentum. To align it with the conventional canonical equations, we can formulate the equation conjugate to Eq. (\ref{11}) as $i\frac{d x_i^*}{dt} = -\frac{\partial H^*}{\partial x_i} = -\frac{\partial H}{\partial x_i} + iQ_i^{(e)}$, where $iQ_i^{(e)} = 2\frac{\partial H_a}{\partial x_i}$, and $H_a = \lim_{N\to\infty}\frac{n\langle\hat{H}_a\rangle}{N}$. Thus, $Q_i^{(e)}$ can be understood as an extra generalized force acting on the system, which is an active force not arising from potential energy.
	
	By substituting Eq. (\ref{2}) into Eq. (\ref{12}), and taking the partial derivative of the classical Hamiltonian function $H$ with respect to $x_i^*$, we obtain the corresponding matrix appearing in Eq. (\ref{11}), as follows:
	\begin{equation}\label{13}
		\begin{split}\begin{aligned}
				\widetilde{H}=h^h+h^a+h^{n}(\vec{x}),
		\end{aligned}\end{split}
	\end{equation}
	with
	\begin{equation}\label{14}
		\begin{split}\begin{aligned}
				h_{i,j}^{n}(\vec{x})=2\sum_{l=1}^{M}\delta_{ij}h_{i,l}\frac{|x_l|^2}{n}-\sum_{k,l=1}^{M}\delta_{ij}h_{k,l}\frac{|x_k|^2|x_l|^2}{n^2}.
		\end{aligned}\end{split}
	\end{equation}	
	Utilizing the properties in Eq. (\ref{4}), we can readily demonstrate that the nonlinear terms in Eq. (\ref{13}) have the following properties: $(h^n)^*(\vec{x}) = (h^n)^{\mathrm{T}}(\vec{x}) = h^n(\vec{x})$.
	
	Next, we will prove the equivalence of Eq. (\ref{10}) and Eq. (\ref{11}). To proceed, we need to express the dynamics of the two-point  function in terms of the time derivatives of the coherent state parameters,
	\begin{equation}\label{15}
		\begin{split}\begin{aligned}
				i\frac{d\sigma_{ij}}{dt}=&i\frac{d}{dt}\Big(\frac{x_i^*x_j}{n}\Big)\\
				=&-\frac{x_j}{n}\Big(i\frac{dx_i}{dt}\Big)^*+\frac{x_i^*}{n}\Big(i\frac{dx_j}{dt}\Big)-\frac{x_i^*x_j}{n^2}\Big(i\frac{dn}{dt}\Big),
		\end{aligned}\end{split}
	\end{equation}					
	where
	\begin{equation}\label{16}
		\begin{split}\begin{aligned}
				i\frac{dn}{dt}=\sum_{k=1}^{M}\Big[x_k^*\Big(i\frac{dx_k}{dt}\Big)-x_k\Big(i\frac{dx_k}{dt}\Big)^*\Big].
		\end{aligned}\end{split}
	\end{equation}
	By utilizing the nonlinear Schrödinger equation (\ref{11}) and its complex conjugate, Eq. (\ref{15})  is transformed as follows: 				
	\begin{equation}\label{17}
		\begin{split}\begin{aligned}				
				i\frac{d\sigma_{ij}}{dt}=&-\frac{x_j}{n}\Big(\sum_{l=1}^{M}\Big[h^h_{i,l}x_l+h^a_{i,l}x_l+h^{n}_{i,l}(\vec{x})x_l\Big]\Big)^*\\
				+&\frac{x_i^*}{n}\Big(\sum_{l=1}^{M}\Big[h^h_{j,l}x_l+h^a_{j,l}x_l+h^{n}_{j,l}(\vec{x})x_l\Big]\Big)\\
				-&\frac{x_i^*x_j}{n^2}\Big(i\frac{dn}{dt}\Big).
		\end{aligned}\end{split}
	\end{equation}	
	Integrating Eqs. (\ref{11}) and (\ref{16}), and utilizing property (\ref{4}), the total probability $n$ (the modulus of the coherent state parameters $\vec{x}$) decays as follows:  
	\begin{align}\label{18}				
		\frac{dn}{dt}=&-i\sum_{k=1}^{M}\Big[x_k^*\Big(\sum_{l=1}^{M}\Big[h^h_{k,l}x_l+h^a_{k,l}x_l+h^{n}_{k,l}(\vec{x})x_l\Big]\Big)\nonumber\\
		-&x_k\Big(\sum_{l=1}^{M}\Big[h^h_{k,l}x_l+h^a_{k,l}x_l+h^{n}_{k,l}(\vec{x})x_l\Big]\Big)^*\Big]\nonumber\\
		=&-i\sum_{k,l=1}^{M}\Big(2h^a_{k,l}x_k^*x_l\Big).
	\end{align}
	From Eqs.~(\ref{17}) and (\ref{18}), using the property $(h^n)^*(\vec{x}) = (h^n)^{\mathrm{T}}(\vec{x}) = h^n(\vec{x})$, and the equality	
	\begin{align}
		&\sum_{l=1}^{M}\Big[h^{n}_{j,l}(\vec{x})\frac{x_i^*x_l}{n}-h^{n}_{i,l}(\vec{x})\frac{x_l^*x_j}{n}\Big]\nonumber\\
		&=2\sum_{k=1}^{M}h_{j,k}\frac{|x_k|^2}{n}\frac{x_i^*x_j}{n}-2\sum_{k=1}^{M}h_{i,k}\frac{|x_k|^2}{n}\frac{x_i^*x_j}{n},
	\end{align}
	
	we arrive at
	\begin{align}\label{20}				
		i\frac{d\sigma_{ij}}{dt}=&-\Big(\sum_{l=1}^{M}\Big[h^h_{l,i}\frac{x_l^*x_j}{n}-h^a_{l,i}\frac{x_l^*x_j}{n}\Big]\Big)+\Big(\sum_{l=1}^{M}\Big[h^h_{j,l}\frac{x_i^*x_l}{n}+h^a_{j,l}\frac{x_i^*x_l}{n}\Big]\Big)\nonumber\\
		+&2\sum_{k=1}^{M}h_{j,k}\frac{|x_k|^2}{n}\frac{x_i^*x_j}{n}-2\sum_{k=1}^{M}h_{i,k}\frac{|x_k|^2}{n}\frac{x_i^*x_j}{n}\nonumber\\
		-&\frac{x_i^*x_j}{n}\sum_{k,l=1}^{M}\Big(2h^a_{k,l}\frac{x_k^*x_l}{n}\Big).
	\end{align}	
	Given the definition $\sigma_{ij} = {x_i^* x_j}/{n}$, a comparison of Eq. (\ref{20}) with Eq. (\ref{10}) readily reveals their equivalence. Since Eq. (\ref{20}) is derived from the nonlinear Schr\"odinger equation (\ref{11}) solely with the aid of the definition of the two-point function, this completes our proof of the equivalence of Eq. (\ref{10}) and Eq. (\ref{11}).
	
	For our specified model (\ref{1}), the coefficient matrix that appears in the extended model (\ref{2}) is explicitly given by
	\begin{align}\label{21}
		\begin{split}		
			&h^h=\begin{pmatrix}0&\Omega&-Je^{-i\pi\gamma}&0\\\Omega&0&0&-Je^{i\pi\gamma}\\-Je^{i\pi\gamma}&0&0&\Omega\\0&-Je^{-i\pi\gamma}&\Omega&0\end{pmatrix},\\
			&h^a=\begin{pmatrix}0&0&0&0\\0&0&0&0\\0&0&-i\beta&0\\0&0&0&-i\beta\end{pmatrix},h=\frac{g}{2}\begin{pmatrix}1&1&0&0\\1&1&0&0\\0&0&1&1\\0&0&1&1\end{pmatrix}.
		\end{split}
	\end{align}
	It is straightforward to verify that the three explicit matrices in Eq. (\ref{21}) satisfy property (\ref{4}). According to Eq. (\ref{12}), the classical Hamiltonian function corresponding to our model (\ref{1}) is given by 
	\begin{align}\label{22}
		H&=\lim_{N\to\infty}\frac{n\langle\hat{H}\rangle}{N} \nonumber\\&=-J(e^{-i\pi x}x_{1}^{*}x_{3}+e^{i\pi x}x_{2}^{*}x_{4}+h.c.)+\Omega(x_{1}^{*}x_{2}+x_{3}^{*}x_{4}+h.c.)\nonumber\\&+\frac{g}{2n}(|x_{1}|^{4}+|x_{2}|^{4}+|x_{3}|^{4}+|x_{4}|^{4})+\frac{g}{n}(|x_{1}|^{2}|x_{2}|^{2}+|x_{3}|^{2}|x_{4}|^{2})\nonumber\\&-i\beta(|x_{3}|^{2}+|x_{4}|^{2}),
	\end{align}
	with $n=|x_{1}|^{2}+|x_{2}|^{2}+|x_{3}|^{2}+|x_{4}|^{2}$. 
	Accoding to Eq. (\ref{18}), the total probability $n$ decays as 
	\begin{align}
		\frac{dn}{dt}=-2\beta(|x_{3}|^{2}+|x_{4}|^{2}).
	\end{align}
	Using Eq. (\ref{11}) in conjunction with the classical Hamiltonian function specified in Eq. (\ref{22}), the components of the unnormalized wave function, represented by the vector $|\varphi\rangle = (x_1, x_2, x_3, x_4)^{\mathrm{T}}$, adhere to the discrete non-Hermitian nonlinear Schr\"odinger equation
	\begin{equation}\label{24}
		\begin{split}\begin{aligned}
				i\frac d{dt}|\varphi\rangle=\widetilde{H}|\varphi\rangle,
		\end{aligned}\end{split}
	\end{equation}
	where
	\begin{equation}\label{25}
		\begin{split}\begin{aligned}
				&\widetilde{H}=\begin{pmatrix}\xi_1&\Omega&-Je^{-i\pi\gamma}&0\\\Omega&\xi_2&0&-Je^{i\pi\gamma}\\-Je^{i\pi\gamma}&0&\xi_3&\Omega\\0&-Je^{-i\pi\gamma}&\Omega&\xi_4\end{pmatrix},
		\end{aligned}\end{split}
	\end{equation}
	with
	\begin{align}
		&\xi_1=\xi_2=-\frac g2\Big[\frac{(|x_1|^2+|x_2|^2)^2}{n^2}+\frac{(|x_3|^2+|x_4|^2)^2}{n^2}\Big]+\frac gn\Big(|x_1|^2+|x_2|^2\Big),\nonumber\\
		&\xi_{3}=\xi_{4}=-\frac{g}{2}\Big[\frac{(|x_{1}|^{2}+|x_{2}|^{2})^{2}}{n^{2}}+\frac{(|x_{3}|^{2}+|x_{4}|^{2})^{2}}{n^{2}}\Big]+\frac{g}{n}\Big(|x_{3}|^{2}+|x_{4}|^{2}\Big)\nonumber\\
		&-i\beta.
	\end{align}
	Apparently, the mean-field Hamiltonian \(\widetilde{H}\) in Eq. (\ref{25}) aligns with the expressions in (\ref{13}) and (\ref{14}) with the  coefficient matrix (\ref{21}) corresponding to our model (\ref{1}).
	
	By gauging away the (irrelevant) global phase via the following transformation:
	\begin{equation}
		|\varphi^{\prime}\rangle=e^{i\theta}|\varphi\rangle,\frac{d\theta}{dt}=-\frac{g}{2}\Big[\frac{(|x_1|^2+|x_2|^2)^2}{n^2}+\frac{(|x_3|^2+|x_4|^2)^2}{n^2}\Big],
	\end{equation}
	Eq. (\ref{24}) becomes
	\begin{equation}	\label{28}
		\begin{split}\begin{aligned}
				i\frac d{dt}|\varphi^{\prime}\rangle=H_m^{\prime}|\varphi^{\prime}\rangle,
		\end{aligned}\end{split}
	\end{equation}
	where
	\begin{equation}	\label{29}
		\begin{split}\begin{aligned}	&H_m'=\begin{pmatrix}\xi_1^{\prime}&\Omega&-Je^{-i\pi\gamma}&0\\\Omega&\xi_2^{\prime}&0&-Je^{i\pi\gamma}&\\-Je^{i\pi\gamma}&0&\xi_3^{\prime}&\Omega\\0&-Je^{-i\pi\gamma}&\Omega&\xi_4^{\prime}\end{pmatrix},
		\end{aligned}\end{split}
	\end{equation}	
	with
	\begin{equation}	
		\begin{split}\begin{aligned}		
				&\xi_{1}^{\prime}=\xi_{2}^{\prime}=\frac gn\Big(|x_1|^2+|x_2|^2\Big),\\
				&\xi_{3}^{\prime}=\xi_{4}^{\prime}=\frac gn\Big(|x_3|^2+|x_4|^2\Big)-i\beta,
		\end{aligned}\end{split}
	\end{equation}
	which forms the basis for our subsequent numerical simulations of the mean-field dynamics. For vanishing dissipation, $\beta=0$,  and the conserved norm $n=1$,  Eq. (\ref{28}) reduces to the Hermitian mean-field model of  SO-coupled bosonic junction, as described in Refs. \cite{60,61,62}.  
	
	\section{Energy spectrum}
	\label{section 3}
	\begin{figure}[htbp]
		\centering
		\includegraphics[width=8cm]{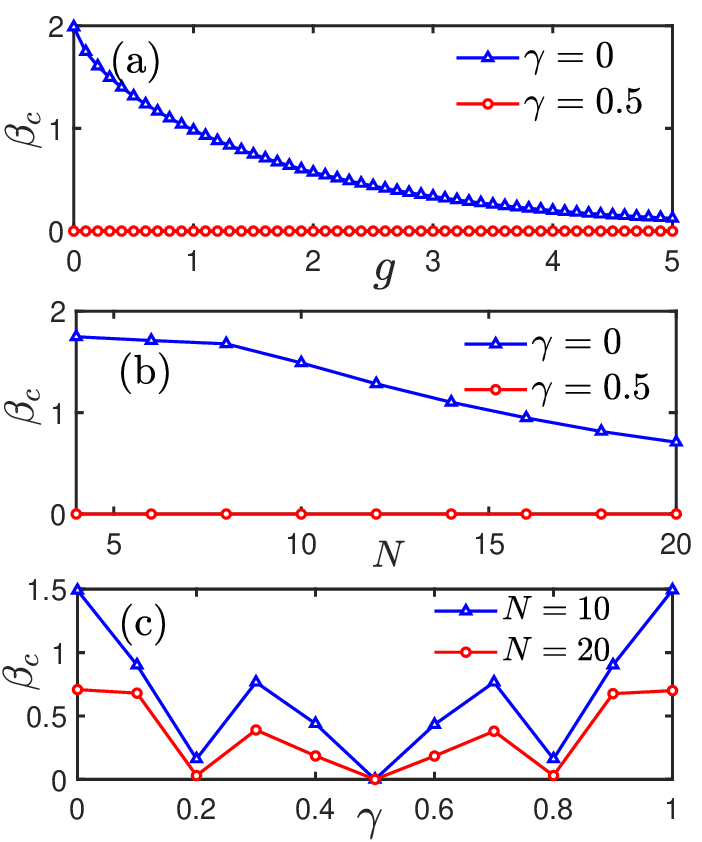}
		\caption{(a) $\mathcal{PT}$-symmetry-breaking threshold \(\beta_c\) for the model (\ref{31}) is shown as a function of the interaction strength \(g\), with SOC strengths of \(\gamma = 0\) and \(\gamma = 0.5\), for \(N=20\). 
			(b) $\mathcal{PT}$-symmetry-breaking threshold \(\beta_c\) as a function of \(N\) at \(\gamma = 0\) and \(\gamma = 0.5\), with a fixed interaction strength \(g=0.1\).
			(c) \(\beta_c\) as a function of SOC strength for \(N = 10\) and \(N = 20\), with a fixed interaction strength \(g=0.1\). Throughout our paper, $J=1$, and all quantities are dimensionless.}\label{fig1}
	\end{figure}
	We first investigate the many-particle spectrum of our model (\ref{1}), which is crucial for the dissipative dynamics. Because the system (\ref{1}) is purely lossy, the imaginary parts of the eigenenergies are all negative, i.e., Im($E$) \textless{} 0, which signifies that the eigenstates are going to decay with time. Due to the normalized physical quantities of interest, we map the system to a $\mathcal{PT}$-symmetric version, as the explicit matrix structure in the Fock basis is essentially equivalent to the original model (\ref{1}). By applying an imaginary energy shift, $\hat{H}\rightarrow\hat{H}+\frac{i\beta\hat{N}}{2}$, the $\mathcal{PT}$-symmetric model is given by
	\begin{equation}\label{31}
		\begin{split}
			\hat{H}_{PT}&=\hat{H}+\frac{i\beta\hat{N}}{2}\\
			&=-J(e^{-i\pi\gamma}\hat{a}_{1}^\dagger\hat{a}_{3}+e^{i\pi\gamma}\hat{a}_{2}^\dagger\hat{a}_{4}+h.c.)+\Omega(\hat{a}_{1}^\dagger\hat{a}_{2}\\&+\hat{a}_{3}^\dagger\hat{a}_{4}+h.c.)+\sum_{i,j=1}^4(\frac{g}{2N}\hat{a}_{i}^\dagger\hat{a}_{j}^\dagger\hat{a}_{i}\hat{a}_{j})\\&-\frac{i\beta}{2}[(\hat{a}_{3}^{\dagger}\hat{a}_{3}+\hat{a}_{4}^{\dagger}\hat{a}_{4})-(\hat{a}_{1}^{\dagger}\hat{a}_{1}+\hat{a}_{2}^{\dagger}\hat{a}_{2})].
		\end{split}
	\end{equation}
	With the definition of the parity operator $\hat{P}$ that interchanges the left and right wells: $\hat{a}_1\rightarrow\hat{a}_3$, $\hat{a}_2\rightarrow\hat{a}_4$, $\hat{a}_3\rightarrow\hat{a}_1$, $\hat{a}_4\rightarrow\hat{a}_2$, and the time-reversal operator $\hat{T}$ as $i\rightarrow -i$, it is evident that the Hamiltonian $\hat{H}_{PT}$ possesses $\mathcal{PT}$ symmetry: $[\hat{P}\hat{T}, \hat{H}_{PT}] = 0$. For $\hat{H}_{PT}$, the energy spectrum $\widetilde{E}$ remains fully real for $\beta$ below the critical value $\beta_c$, and becomes complex (appears in complex-conjugate pairs) for $\beta > \beta_c$. Thus, $\beta_c$ represents a phase transition between the broken and unbroken $\mathcal{PT}$ symmetry regions. Given that the energy spectrum of $\hat{H}_{PT}$ is essentially a shift in the imaginary part of the energy spectrum of $\hat{H}$, i.e., $\widetilde{E} = E + i\beta N/2$, this means that the imaginary parts of the eigenvalues of $\hat{H}$ are symmetrically distributed around $-\beta N/2$.   
	
	In Fig.~\ref{fig1}, we have numerically investigated the dependence of the symmetry-breaking threshold $\beta_c$ on the interaction strength $g$, the particle number $N$, and the SOC strength $\gamma$. A key finding is that when the SOC strength $\gamma = 0.5$, the $\mathcal{PT}$symmetry is fragile to nonzero non-Hermiticity, indicating that the system is in the $\mathcal{PT}$-broken phase for any arbitrarily small value of $\beta$. We also observe that the zero symmetry-breaking threshold $\beta_c$ for $\gamma = 0.5$ is a universal characteristic applicable to any particle number and interaction strength. In Fig. \ref{fig1} (c), we find that the symmetry-breaking threshold $\beta_c$ is very close to zero when $\gamma$ is approximately 0.2 and 0.8 for $N=20$, but this is accidental, as zero $\beta_c$ can be removed by changing the particle number.
	
	To provide some initial physical insights into the key observations presented in Fig.~\ref{fig1}, we first consider the limiting case of zero atomic interaction. We can analytically treat the noninteracting problem by employing the representation of bosonic creation and annihilation operators in generalized Bargmann space \cite{55,56,102}, with the following substitution:
	\begin{equation}
		\begin{split}
			&\hat{a}_{i}^{\dagger} \rightarrow Z_i,\hat{a}_{i} \rightarrow \frac{\partial}{\partial Z_i},\\
			&|\psi\rangle=\sum_{i,j,k,l}f(i,j,k,l)(\hat{a}_{1}^{\dagger})^i(\hat{a}_{2}^{\dagger})^j(\hat{a}_{3})^k(\hat{a}_{4})^l|0\rangle\\
			&\rightarrow \psi(Z_1,Z_2,Z_3,Z_4)=\sum_{i,j,k,l}f(i,j,k,l)Z_1^iZ_2^jZ_3^kZ_4^l.
		\end{split}
	\end{equation}
	In the Bargmann space, the eigenvalue equation $\hat{H}_{PT}|x_1,x_2,x_3,x_4,N\rangle=\widetilde{E}|x_1,x_2,x_3,x_4,N\rangle$ can be expressed as follows:
	\begin{equation}\label{33}
		\begin{split}
			\begin{aligned}
				\hat{H}_{PT}^Bf_{x,N}(Z)=&\widetilde{E}f_{x,N}(Z)\\
				=&N\frac{1}{\sqrt{N!}}\Big[\sum_{i,j=1}^{4}\Big(h^h_{i,j}+h^a_{i,j}+\frac{i\beta}{2}\delta_{i,j}\Big)x_jZ_i\Big]
				\\&\times\Big(x_1Z_1+x_2Z_2+x_3Z_3+x_4Z_4\Big)^{N-1},
			\end{aligned}
		\end{split}
	\end{equation}
	where the $N$-particle coherent state $|x_1,x_2,x_3,x_4,N\rangle$ is represented by the analytical function $f_{x,N}(Z)=\frac{1}{\sqrt{N!}}(x_1Z_1+x_2Z_2+x_3Z_3+x_4Z_4)^N$, and the operator corresponding to $\hat{H}_{PT}$ is represented by $\hat{H}_{PT}^B$ in the Bargmann space. From Eq. (\ref{33}), it follows that   
	\begin{equation}
		\begin{split}
			&(h^h+h^a+\frac{i\beta}{2})(x_1,x_2,x_3,x_4)^{\mathrm{T}}=\varepsilon(x_1,x_2,x_3,x_4)^{\mathrm{T}},\\
			&\hat{H}_{PT}|x_1,x_2,x_3,x_4,N\rangle=N\varepsilon|x_1,x_2,x_3,x_4,N\rangle.
		\end{split}
	\end{equation}
	This is a straightforward physical consequence: in the noninteracting case, if the state vector $|\psi\rangle = (x_1, x_2, x_3, x_4)^{\mathrm{T}}$ is an eigenstate of the single-particle Hamiltonian $h^h+h^a+\frac{i\beta}{2}$, then the corresponding $N$-particle coherent state $|x_1,x_2,x_3,x_4,N\rangle$ is also an eigenstate of the many-particle Hamiltonian $\hat{H}_{PT}$, with its eigenvalues being $N$ times the eigenvalues of the single-particle Hamiltonian. Consequently, insights into the many-particle spectrum can be derived directly from the single-particle Hamiltonian when $g=0$.
	
\begin{figure}[htp]
		\centering
		\includegraphics[width=8cm]{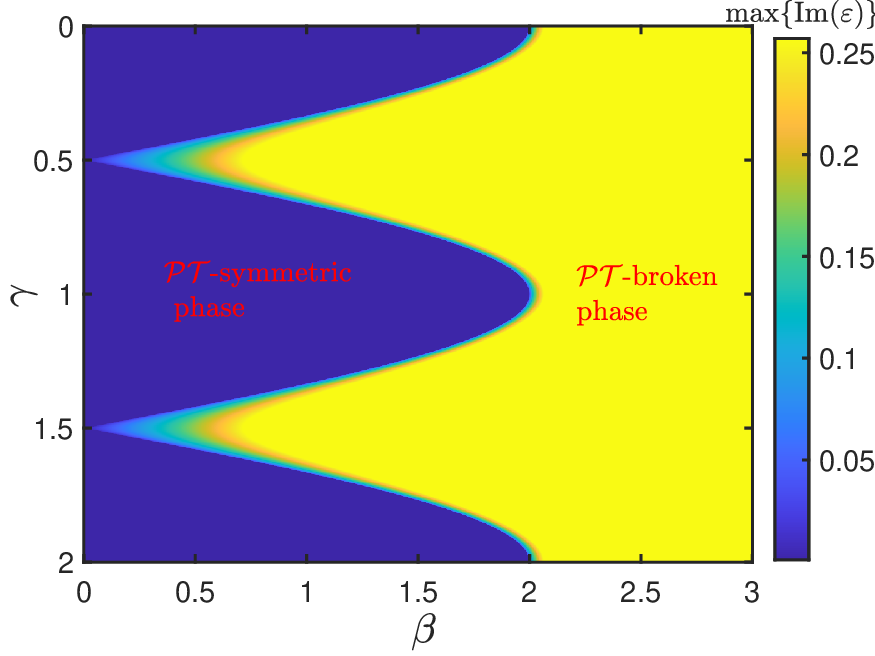}
		\caption{The phase diagram of the \(\mathcal{PT}\)-symmetric single-particle Hamiltonian \(h^h + h^a + \frac{i\beta}{2}\), with $h^h$ and $h^a$ given in Eq.~(\ref{21}). The color map illustrates the different values of the maximum imaginary parts of the eigenvalues $\varepsilon$. The blue area signifies the \(\mathcal{PT}\)-symmetric phase, whereas the yellow area indicates the \(\mathcal{PT}\)-broken phase.}\label{fig2}
	\end{figure}

	In Fig.~\ref{fig2}, we have plotted the $\mathcal{PT}$-symmetric phase diagram for the single-particle Hamiltonian by calculating the maximum imaginary part of the eigenvalues of $h^h+h^a+\frac{i\beta}{2}$ as functions of the SOC strength $\gamma$ and the gain-loss parameter $\beta$. It is observed that the $\mathcal{PT}$-symmetry-breaking phase transition occurs even for arbitrarily small non-Hermitian parameters when the SOC strength assumes a half-integer value. We can view the particle interaction as a perturbation that may enhance $\mathcal{PT}$-symmetry breaking, which partially explains the many-particle phenomenon observed in Fig.~\ref{fig1}, where $\mathcal{PT}$-symmetry breaking occurs for arbitrarily weak non-Hermitian parameters when $\gamma=0.5$.\\
	\begin{figure}[htp]
		\centering
		\includegraphics[width=8cm]{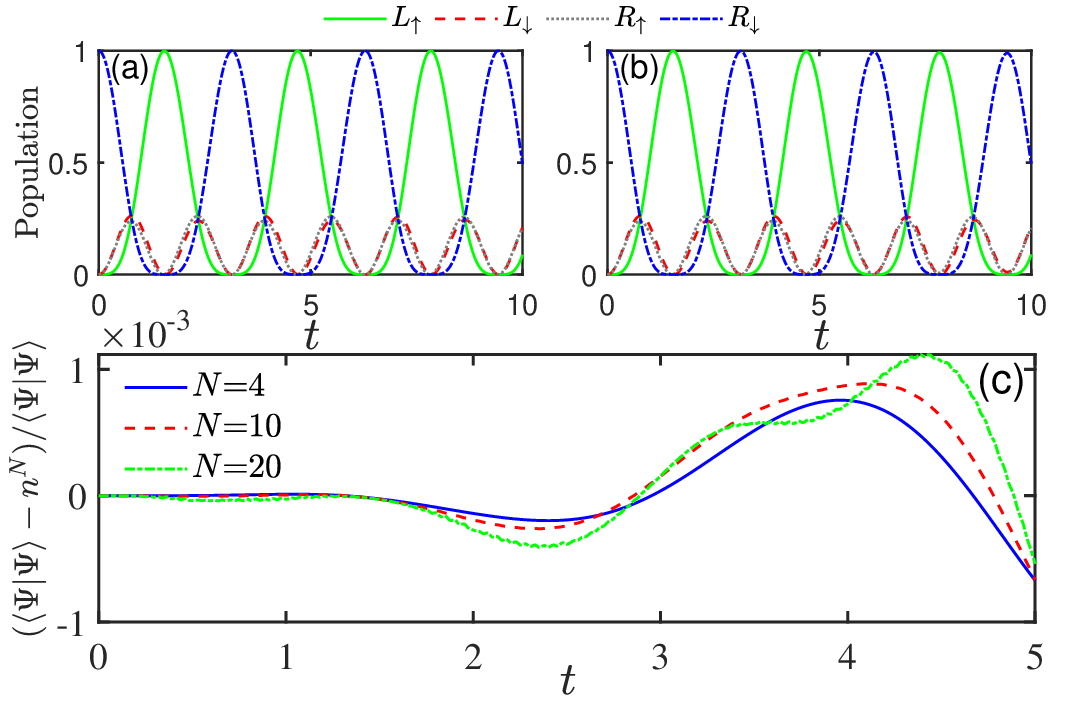}
		\caption{The time evolution of the populations on all four modes is shown: (a) for the full-quantum system (\ref{1}) with \(N=4\), and (b) for the mean-field model (\ref{29}). (c) displays the relative errors between the many-particle and mean-field results for particle numbers \(N=4\), \(10\), and \(20\). Initially, the entire BEC is localized in the right well, fully occupying the spin-down state. The other parameters are set to \(g=0.1\), \(\gamma=0\), and \(\beta=0.1\).}\label{fig3}
	\end{figure}
	
	\section{Mean-field and many-particle dynamics }
	Let us now make a comparison between the mean-field evolution and the many-particle dynamics.  The mean-field evolution is obtained by integrating the nonlinear Schr\"odinger equation (\ref{28}) for a state initially prepared as $|\psi(0)\rangle=(0, 0, 0, 1)^{\mathrm{T}}$, while the full quantum solution is obtained by numerically simulating the Bose-Hubbard Hamiltonian (\ref{1}) using the corresponding initial coherent state with unit norm.  For a better comparison, we directly depict  the dynamics of the populations across all four modes, finding that even with a small number of particles ($N=4$), the mean-field dynamics  [Fig.~\ref{fig3}(a)] agrees well with the quantum evolution  [Fig.~\ref{fig3}(b)]  over a short time scale. This observation is confirmed by the relative deviations between the mean-field probability $n(t)$ and the normalization of the many-particle wave function, as shown in Fig.~\ref{fig3}(c).
	
	To delve deeper into the correspondence between many-particle and mean-field dynamics, we focus on two key physical quantities: the atomic population imbalance $Z$ and the  spin population imbalance $I$ between two wells, with their time derivatives yielding the atomic current and the net spin current, respectively. Defining the operator $\hat{Z} := \hat{a}_{L\uparrow}^\dagger\hat{a}_{L\uparrow} + \hat{a}_{L\downarrow}^\dagger\hat{a}_{L\downarrow} - \hat{a}_{R\uparrow}^\dagger\hat{a}_{R\uparrow} - \hat{a}_{R\downarrow}^\dagger\hat{a}_{R\downarrow}$ that represents the particle population difference between two wells, and the operator $\hat{I} := \hat{a}_{L\uparrow}^\dagger\hat{a}_{L\uparrow} - \hat{a}_{L\downarrow}^\dagger\hat{a}_{L\downarrow} - \hat{a}_{R\uparrow}^\dagger\hat{a}_{R\uparrow} + \hat{a}_{R\downarrow}^\dagger\hat{a}_{R\downarrow}$ that represents the  spin population difference between two wells, the corresponding expectation values give rise to
	\begin{equation}\label{35}
		\begin{split}\begin{aligned}
				Z=&\langle\hat{Z}\rangle=\langle\hat{a}_{L\uparrow}^\dagger\hat{a}_{L\uparrow}\rangle+\langle\hat{a}_{L\downarrow}^\dagger\hat{a}_{L\downarrow}\rangle-\langle\hat{a}_{R\uparrow}^\dagger\hat{a}_{R\uparrow}^\dagger\rangle-\langle\hat{a}_{R\downarrow}^\dagger\hat{a}_{R\downarrow}\rangle,\\
				I=&\langle\hat{I}\rangle=\langle\hat{a}_{L\uparrow}^\dagger\hat{a}_{L\uparrow}\rangle-\langle\hat{a}_{L\downarrow}^\dagger\hat{a}_{L\downarrow}\rangle-\langle\hat{a}_{R\uparrow}^\dagger\hat{a}_{R\uparrow}^\dagger\rangle+\langle\hat{a}_{R\downarrow}^\dagger\hat{a}_{R\downarrow}\rangle. 
		\end{aligned}\end{split}
	\end{equation}
	In the aforementioned definition of the corresponding expectation values of operators, we have omitted the contribution from the variation in the norm, such that the dynamics of the physical quantities are the same for both the original model (\ref{1}) and the $\mathcal{PT}$-symmetric model (\ref{31}). In the following, we shall refer to the system as being in either the $\mathcal{PT}$ phase or the $\mathcal{PT}$-broken phase, depending on the eigenstate properties of the $\mathcal{PT}$-symmetric model. In the $\mathcal{PT}$-broken regime, the purely decaying system (\ref{1}) eventually evolves into the eigenstate with the largest imaginary eigenvalue after a sufficient amount of time, which we refer to as the stable state or the surviving state.
	\begin{figure}[htp]
		\centering
		\includegraphics[width=8cm]{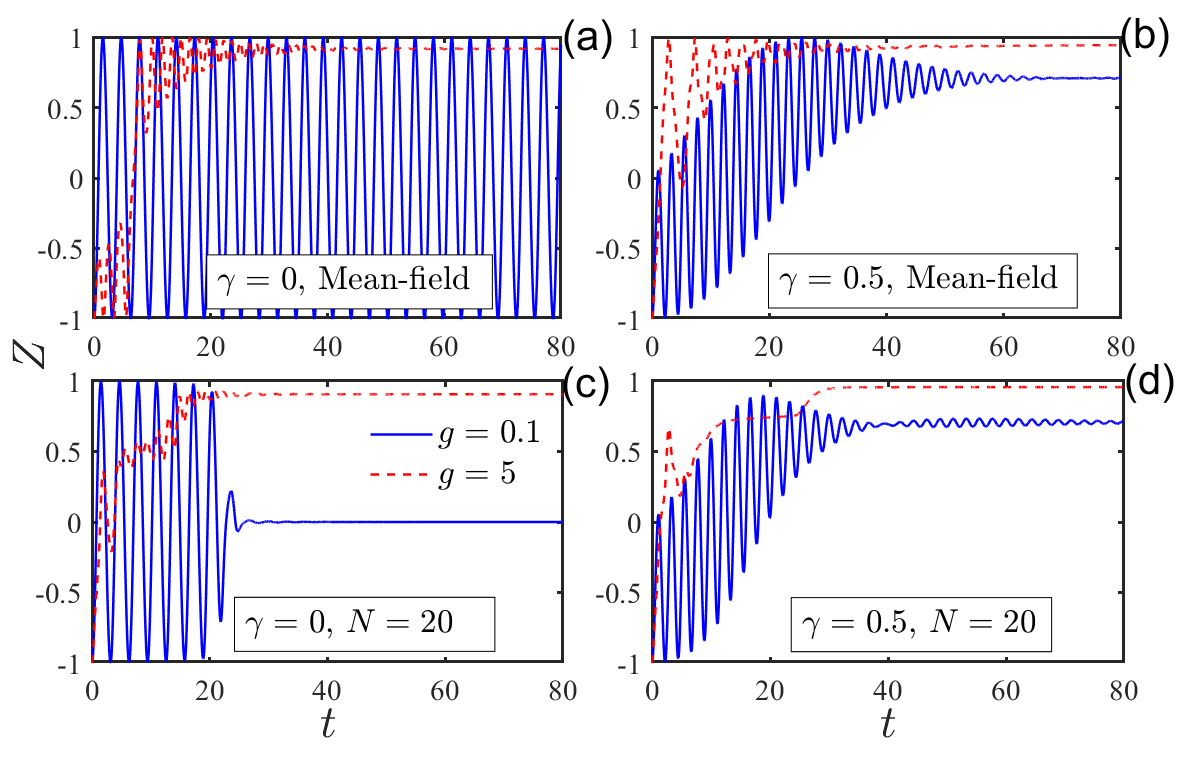}
		
		\caption{The time evolution of the particle population difference \(Z\) between the two wells, defined by Eq. (\ref{35}), for \(\gamma = 0\) (left column) and \(\gamma = 0.5\) (right column). (a) and (b) illustrate the evolution under the mean-field scenario, whereas (c) and (d) depict the evolution for full-quatum system with \(N = 20\). In all panels, solid lines represent the \(g = 0.1\) case, while dashed lines correspond to the \(g = 5\) case. The initial state is the same as in Fig.~\ref{fig3}.}\label{fig4}
	\end{figure}
	
\begin{figure}[htp]
		\centering
		\includegraphics[width=8cm]{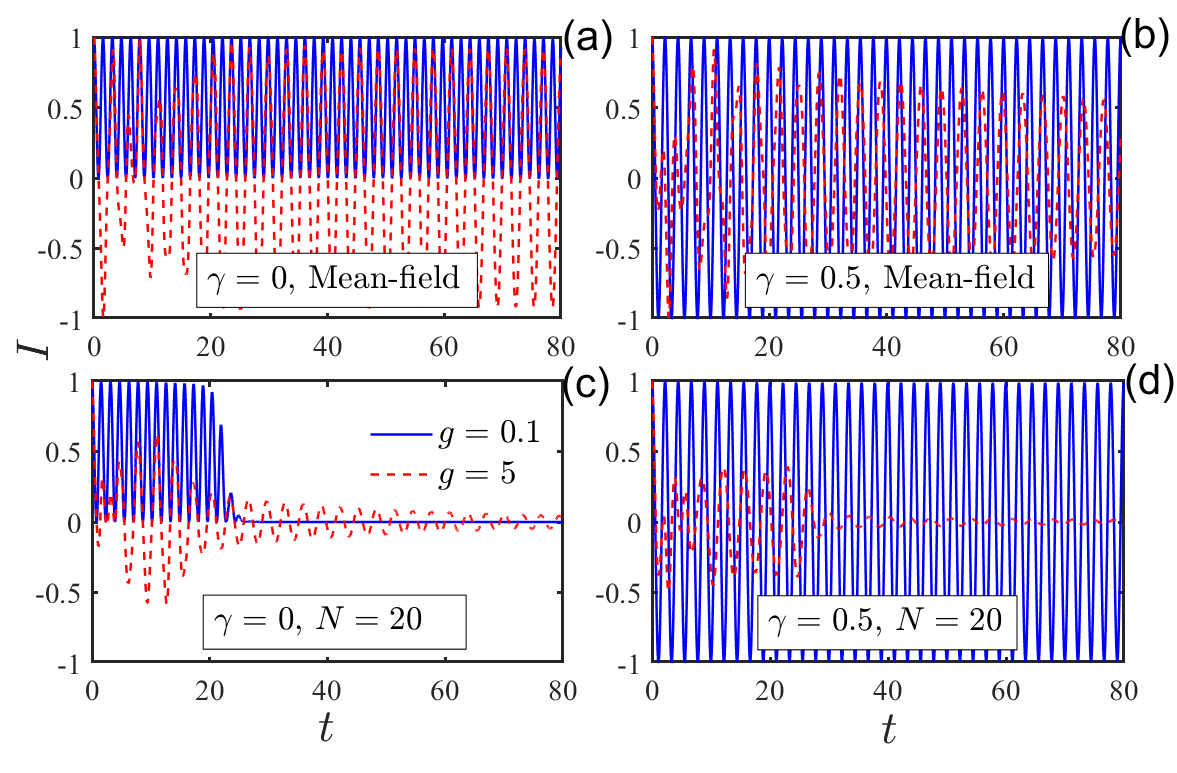}
		\caption{The time evolution of \(I\), representing the  spin population difference between the two wells as defined by Eq. (\ref{35}), is shown for system parameters and initial states that are exactly the same as those in Fig.~\ref{fig4}.}\label{fig5}
	\end{figure}

	In Fig.~\ref{fig4}, we show the time evolution of the particle population difference $Z$ between two wells for a system initialized with the spin-down state in the right well. The left column displays the case for $\gamma=0$, whereas the right column shows the case for $\gamma=0.5$. In the case of $\gamma = 0$, when the interaction strength and non-Hermiticity are small ($g = 0.1$ and $\beta = 0.1$), the classical mean-field dynamics exhibit Rabi-like oscillations. However, in the many-particle system, we observe the familiar breakdown behavior where the many-particle motion oscillates with a decreasing amplitude until it is damped to zero. As the interaction strength is increased to $g=5$, both the mean-field and full quantum calculations show that the particle population difference $Z$ tends to settle at a positive steady value, indicating that the corresponding $\mathcal{PT}$-symmetric system is in the $\mathcal{PT}$-broken regime. This result is consistent with that shown in Fig.~\ref{fig1}, where we observe that the symmetry-breaking threshold, which distinguishes between the $\mathcal{PT}$-symmetric and $\mathcal{PT}$-broken phases, decreases with increasing interaction strength. For $\gamma = 0.5$, as discussed earlier, the system is always in the $\mathcal{PT}$-broken phase, and thus the mean-field and many-particle population imbalances $Z$ are seen to tend to nonzero constant values over time for any set of parameters. From Fig.~\ref{fig4}, we clearly see a tendency that the larger the particle interaction strength, the larger the steady value that the particle population difference $Z$ eventually settles into. With the same initial condition and parameter set as used in Fig.~\ref{fig4}, the time evolution of the spin population difference $I$ is presented in Fig.~\ref{fig5}, as computed using both the mean-field and full quantum treatments. In the $\mathcal{PT}$-broken regime, a peculiar quantum phenomenon observed in Fig.~\ref{fig5} is that, as time elapses, the particle population difference tends towards a nonzero constant value, whereas the  spin population difference exhibits a periodic oscillatory pattern.
	\begin{figure}[htp]
		\centering
		\includegraphics[width=8cm]{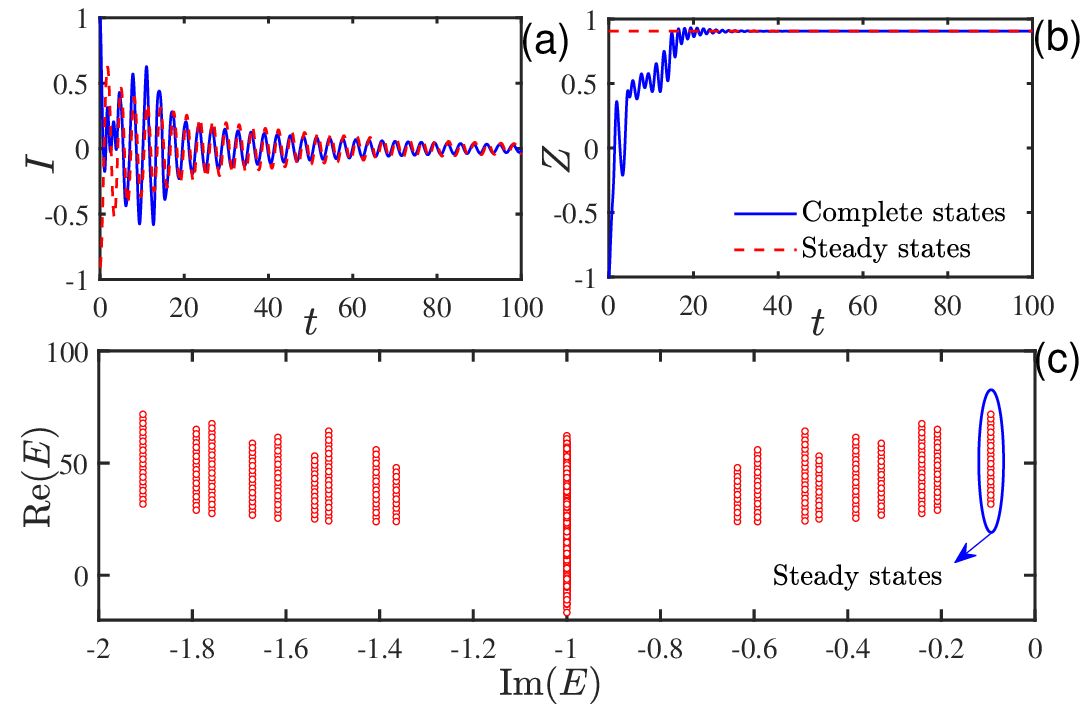}
		
		\caption{Many-particle dynamics for \(I\) (a) and \(Z\) (b) with \(N=20\). The solid blue lines represent results calculated using the complete state basis, which are in good agreement with the results (red dashed line) from Eq. (\ref{38}) utilizing the eigenstates with largest imaginary parts  [indicated by blue circles in (c)] within the steady-state subspace. (c) The energy spectrum of model (\ref{1}) in the complex plane. The system parameters are \(\beta = 0.1\), \(g = 5\), \(\gamma = 0\), and the initial states in (a) and (b) are the same as those in Fig.~\ref{fig4}.}\label{fig6}
	\end{figure}
	
	We further understand the physics presented in Figs.~\ref{fig4} and \ref{fig5} by carefully examining the eigenstates and eigenvalues of model (\ref{1}). Taking the parameter set $\gamma=0$, $\beta=0.1$, $g=5$ as an example, we find that there exist multiple eigenstates corresponding approximately to the same largest imaginary parts of the eigenvalues [see Fig.~\ref{fig6}(c)]. The fact that there is degeneracy in the largest imaginary eigenvalues (that is, multiple stable states exist) is a general feature in the model we are considering. As a general case, we arrange the eigenvalues of model (\ref{1}) in descending order according to the imaginary part of the eigenvalues as follows:  
	\begin{equation}
		\begin{split}\begin{aligned}
				&\hat{H}|n\rangle=E_n|n\rangle,\\
				&0\textgreater\mathrm{Im}(E_1)\approx\cdots\approx \mathrm{Im}(E_l)\textgreater \mathrm{Im}(E_{l+1})\geqslant\cdots\geqslant \mathrm{Im}(E_D),
		\end{aligned}\end{split}
	\end{equation}
	where $D=C_{N+3}^3$ is the dimension of Hilbert space for $N$-particle system. We refer to the Hilbert space composed of the eigenstates $|1\rangle, |2\rangle, \ldots, |l\rangle$, which share the same largest imaginary part of the eigenvalues, as a steady-state (SS) subspace. At the initial time, the quantum state can be expanded as $|\Psi(0)\rangle = \sum_{n=1}^{D} C_{n} |n\rangle$. With time evolution, the quantum state evolves according to
	\begin{equation}
		\begin{split}\begin{aligned}
				|\Psi(t)\rangle=&\sum_{n=1}^{D}e^{-iE_nt}C_{n}|n\rangle=\sum_{n=1}^{D}e^{-i\mathrm{Re}(E_{n})t}e^{\mathrm{Im}(E_{n})t}C_{n}|n\rangle,\\
				\overset{t\to\infty}{=}&\sum_{n=1}^{l}e^{-i\mathrm{Re}(E_{n})t}e^{\mathrm{Im}(E_{n})t}C_{n}|n\rangle. 
		\end{aligned}\end{split}
	\end{equation}
	After a long period of time, the surviving state is given by $|\Psi(t)\rangle_s = \sum_{n=1}^{l} e^{-i\mathrm{Re}(E_{n})t} e^{\mathrm{Im}(E_{n})t} C_{n} |n\rangle$. We define the atomic population difference \( Z_s \) and  spin population difference \( I_s \) in terms of the surviving state in the steady-state subspace as follows
	\begin{equation}\label{38}
		\begin{split}\begin{aligned}
				Z_s=&\frac{_s\langle\Psi|\hat{Z}|\Psi\rangle_s}{_s\langle\Psi|\Psi\rangle_s},\qquad I_s=&\frac{_s\langle\Psi|\hat{I}|\Psi\rangle_s}{_s\langle\Psi|\Psi\rangle_s}. 
		\end{aligned}\end{split}
	\end{equation}
	
	We consider the simple case where $l=1$, which means there is no degeneracy in the largest imaginary part of the eigenvalues and only one eigenstate survives. Then, we have $|\Psi(t)\rangle_s = e^{-iE_{1}t}C_{1}|1\rangle$, leading to $Z_s = \frac{\langle1|\hat{Z}|1\rangle}{\langle1|1\rangle}$ and $I_s = \frac{\langle1|\hat{I}|1\rangle}{\langle1|1\rangle}$, both of which are independent of time. Consequently, both $Z$ and $I$ asymptotically tend to a constant value. For $l > 1$, since the real parts of the $l$ eigenvalues in the steady-state subspace are usually not the same, even though the imaginary parts are equal, it was generally expected that both $I$ and $S$ should exhibit oscillatory patterns. However, from Figs.~\ref{fig4} and \ref{fig5}, we find that in the $\mathcal{PT}$-broken regime, after a long period of time, $Z$ tends to a nonzero constant value, while $I$ exhibits oscillatory patterns. For example, see the blue lines in the right column of Figs.~\ref{fig4} and \ref{fig5}. In Figs.~\ref{fig6}(a) and \ref{fig6}(b), with the specific parameter $\beta = 0.1$, $g = 5$, $\gamma = 0$, and $N = 20$, we present the time evolutions of $Z$ and $I$ using the complete eigenstate basis (blue solid lines) and the eigenstates [marked in Fig.~\ref{fig6}(c)] in the steady-state subspace (red dashed lines), respectively. We observe that both $I$ and $I_s$ [given by Eq. (\ref{38})] display oscillatory behavior, and after an initial transient time, they match each other well, whereas from the beginning of time, $Z_s$ [given by Eq. (\ref{38})] remains a constant value, which coincides with the one that the system eventually settles into. The oscillation of $I$ while $Z$ does not oscillate may be due to the fact that, in Eq. (\ref{38}), the time-dependent terms for $Z$ cancel each other out exactly, whereas for $I$, these terms do not. This situation was also revealed  in Ref. \cite{59}. 
	\begin{figure}[htp]
		\centering
		\includegraphics[width=8cm]{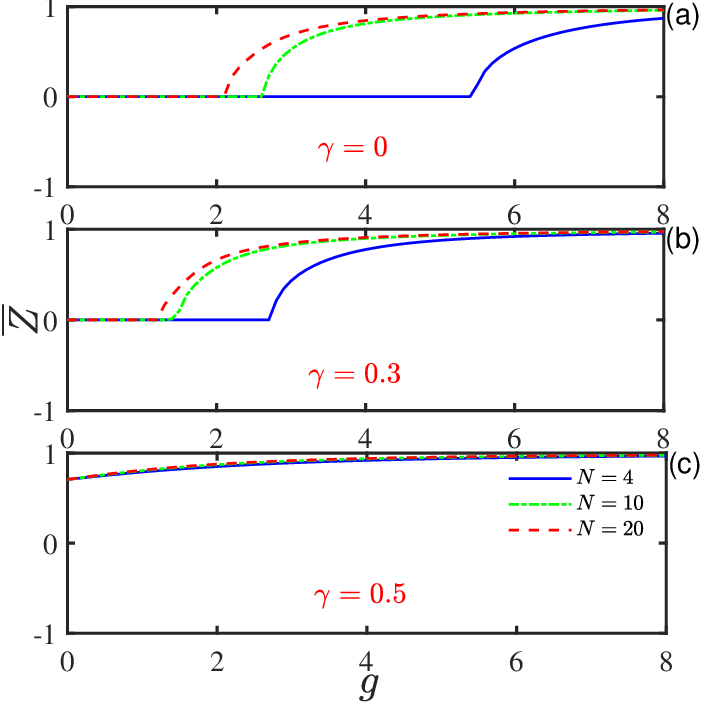}
		
		\caption{Time-averaged  particle population imbalance, \(\overline{Z}\), as a function of the interaction strength \(g\), is shown for a full-quantum system with different particle numbers: \(N = 4\), \(N = 10\), and \(N = 20\). From top to bottom, the plots correspond to \(\gamma = 0\), \(\gamma = 0.3\), and \(\gamma = 0.5\). The initial state and other system parameters are the same as those in Fig.~\ref{fig4}.}\label{fig7}
	\end{figure}
	
	In Fig.~\ref{fig7}, we plot the time-averaged atomic population imbalance, defined as $\overline{Z} = \lim_{t\to\infty} \frac{1}{t} \int_{0}^{t} Z(\tau) d\tau$, against the interaction strength $g$ for model (\ref{1}) with different values of $N$, using the same initial conditions as in Fig.~\ref{fig4}. As shown in Figs.~\ref{fig7}(a) and (b), for $\gamma=0$ and $\gamma=0.3$, there exists a symmetry-breaking phase transition, where $\overline{Z}$ changes from zero to nonzero values when $g$ exceeds a critical threshold. For the interaction strength $g$ greater than the phase transition point, the particles become localized in the left well without any loss, and hence the average of population difference between the two wells will not be zero and will increase as the interaction strength increases. From Figs. \ref{fig7}(a) and (b), it is clearly seen that with an increasing number of particles, the transition point shifts to the left. This means that the symmetry-breaking phase transition shows a strong dependence on the particle number $N$. For $\gamma = 0.5$, as the system is always in the $\mathcal{PT}$-broken phase, the dependence of $\overline{Z}$ on $g$ appears to be less sensitive to the particle number, as shown in Fig. \ref{fig7}(c). In this case, the phenomenon of atomic self-trapping (with nonzero $\overline{Z}$) in the lossless well occurs for any particle number $N$ and any interaction strength $g$.
	\begin{figure}[htp]
		\centering
		\includegraphics[width=8cm]{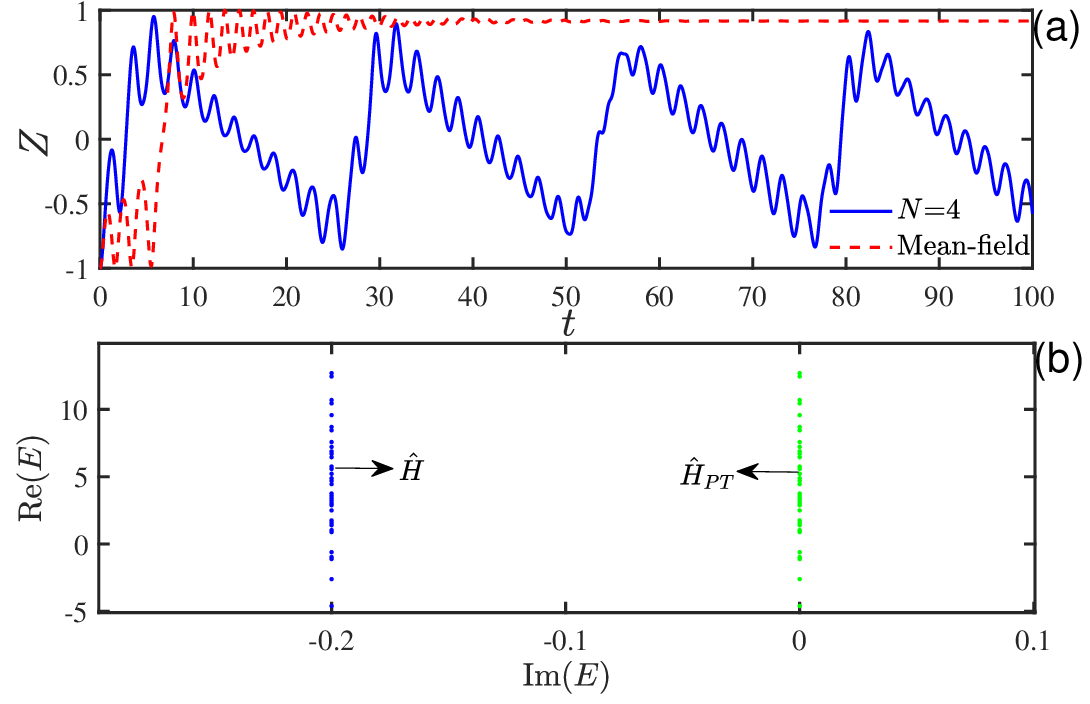}
		\caption{(a) Breakdown of the correspondence between the quantum dynamics (the blue solid line) and the mean-field dynamics (the red dashed line). (b) The energy spectrum in the complex plane for $N = 4$. The spectrum, derived from model (\ref{1}), corresponds to an entirely real spectrum associated with the $\mathcal{PT}$-symmetric Hamiltonian $\hat{H}_{PT}$ after applying an imaginary energy shift. The time evolution of $Z$ shown in (a) is given under the same initial condition as used in Fig.~\ref{fig3}. The parameters are $\beta = 0.1$, $g = 5$, $\gamma = 0$, and $N = 4$. }\label{fig8}
	\end{figure}
	
	From Fig.~\ref{fig7}, we can expect that for $\gamma=0$, the deviation from the mean-field critical point is inversely proportional to $N$, and thus the mean-field phase transition point deviates significantly from that of the full-quantum system with relatively small $N$. As shown in Fig.~\ref{fig3}, for a small system size with $N=4$ and at a small interaction strength $g=0.1$, the mean-field dynamics closely matches the quantum evolution during certain initial time intervals. This is because the parameter is below the critical phase transition point for both mean-field and full quantum systems. When the parameters lie in the vicinity of the phase transition point of few-particle quantum system, for example, at $g=5$, quantum fluctuations become huge, and mean-field dynamics and quantum dynamics can display qualitatively different results. As shown in Fig.~\ref{fig8}(a), with $\beta=0.1$, $g=5$, $\gamma=0$, and $N=4$, and using the same initial conditions as in Figs.~\ref{fig3} and \ref{fig4}, we can observe that the atomic population difference $Z$ oscillates for a considerable amount of time in the full quantum system, whereas the corresponding mean-field atomic population difference tends to a constant value. This means that, in general, near the full-quantum phase transition point, quantum fluctuations become significant, making the correspondence between the classical and quantum dynamics more likely to break down. Through examination of the full-quantum spectrum as shown in Fig. \ref{fig8}(b), we can intuitively understand the breakdown of the mean-field and quantum correspondence. It is evident that the imaginary part of the energy spectrum of model (\ref{1}) in this case is nearly degenerate, which corresponds to an entirely real spectrum of the corresponding $\mathcal{PT}$-symmetric Hamiltonian (\ref{31}), indicating that the full-quantum system is in the $\mathcal{PT}$-symmetric phase. In contrast, for such a large interaction strength, the mean-field dynamics obviously resides in the $\mathcal{PT}$-broken phase.
	\begin{figure}[htp]
		\centering
		\includegraphics[width=8cm]{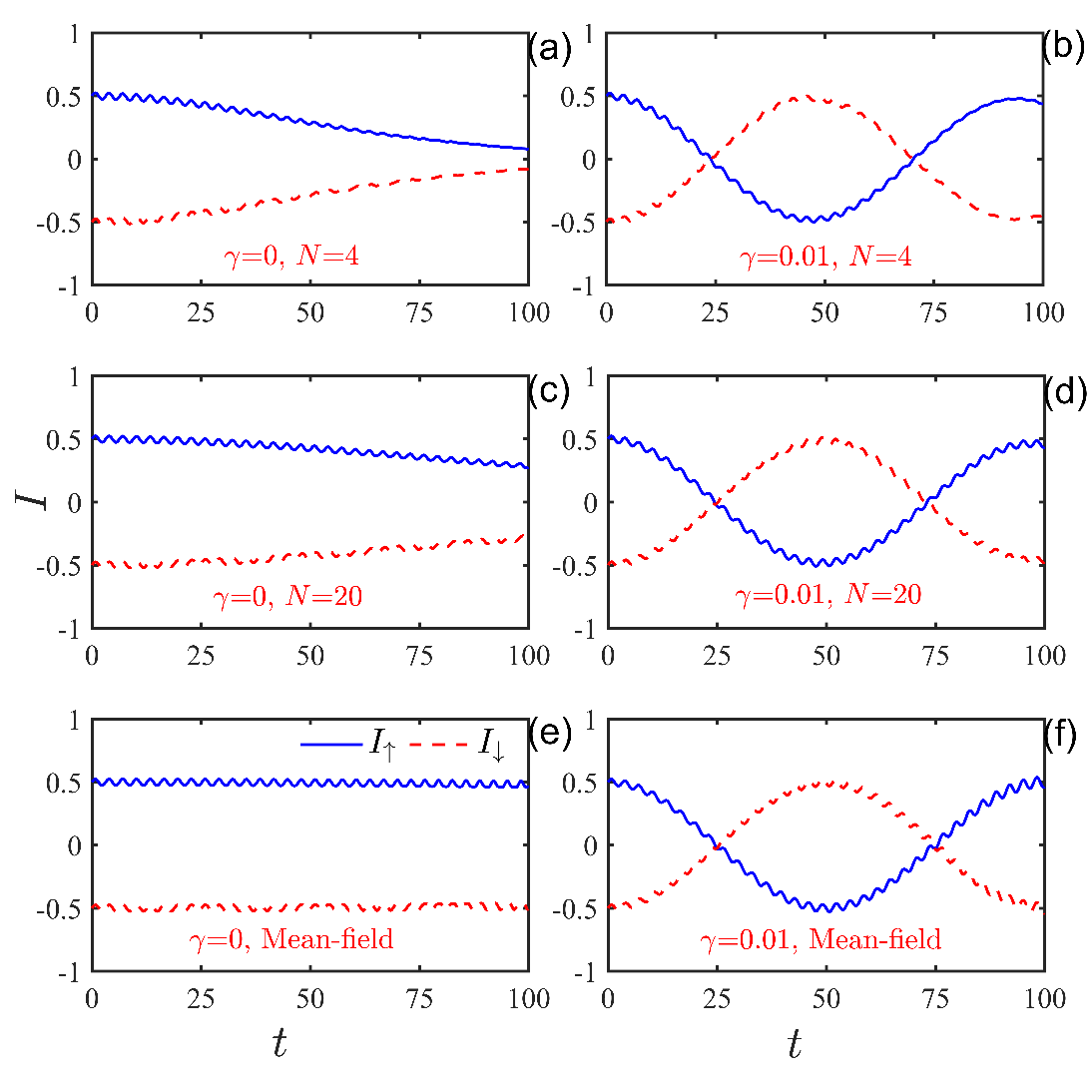}
		\caption{The time evolution of  $I_\uparrow=\langle\hat{a}_{L\uparrow}^\dagger\hat{a}_{L\uparrow}\rangle-\langle\hat{a}_{R\uparrow}^\dagger\hat{a}_{R\uparrow}^\dagger\rangle$ (blue solid lines) and $I_\downarrow =\langle\hat{a}_{L\downarrow}^\dagger\hat{a}_{L\downarrow}\rangle-\langle\hat{a}_{R\downarrow}^\dagger\hat{a}_{R\downarrow}\rangle$ (red dahed lines) for the two spin compoents  for  \( \gamma = 0 \) (left column) and  \( \gamma = 0.01 \) (right column), with $\beta=0.1$. From top to bottom, the plots correspond to \( N = 4 \), \( N = 20 \), and mean-field dynamics. To study the synchronization dynamics, the initial state is prepared as $|\psi(0)\rangle=\frac{1}{\sqrt{2}}(1, 0, 0, 1)^{\mathrm{T}}$ for mean-field dynamics, and as $|\Psi(0)\rangle=\frac{1}{\sqrt{N!}} \left(\frac{1}{\sqrt{2}}\hat{a}_{L\uparrow}^\dagger+\frac{1}{\sqrt{2}}\hat{a}_{R\downarrow}^\dagger\right)^N |0\rangle$ for the full-quantum dynamics.}\label{fig9}
	\end{figure}
	
	Next, we proceed to compare the mean-field and quantum dynamics more deeply by investigating the so-called measure synchronization dynamics of the spin-up and spin-down components within the dissipative system, which is useful for establishing functional quantum correlations \cite{58}. The measure synchronization behaviors have been extensively studied in the two-species bosonic-Josephson junction system\cite{94,95}. Upon its occurrence, the orbits of the spin-up $\ket{\uparrow}$ and spin-down $\ket{\downarrow}$ subsystems cover the same region of phase space, which is related to the energy exchange between the two subsystems\cite{95}. In our model, the control parameters that couple the two spin states are the Raman coupling and the particle interactions between different spin (hyperfine) states. To study the measure synchronization behaviors, we initialized the system in state $|\psi(0)\rangle=\frac{1}{\sqrt{2}}(1, 0, 0, 1)^{\mathrm{T}}$ for the mean-field dynamics and the corresponding coherent state $|\Psi(0)\rangle=\frac{1}{\sqrt{N!}} \left(\frac{1}{\sqrt{2}}\hat{a}_{L\uparrow}^\dagger+\frac{1}{\sqrt{2}}\hat{a}_{R\downarrow}^\dagger\right)^N |0\rangle$, for the many-particle dynamics. That is, all particles are initially prepared with equal numbers of spin-up and spin-down states. In the following numerical investigations, we use $I_\uparrow$ and $I_\downarrow$ to represent the motion of the two spin components: $I_\uparrow=\langle\hat{a}_{L\uparrow}^\dagger\hat{a}_{L\uparrow}\rangle-\langle\hat{a}_{R\uparrow}^\dagger\hat{a}_{R\uparrow}\rangle $, $I_\downarrow=\langle\hat{a}_{L\downarrow}^\dagger\hat{a}_{L\downarrow}\rangle-\langle\hat{a}_{R\downarrow}^\dagger\hat{a}_{R\downarrow}\rangle $.
	
	In Fig.~\ref{fig9}, we show the time evolution of $I_\sigma$ for the spin-up ($\sigma=\uparrow$) and spin-down ($\sigma=\downarrow$) states, with the fixed parameters \( g = 0.1\), \(\beta=0.1 \), for $\gamma=0$ (left column) and $\gamma=0.01$ (right column), respectively. For $\gamma=0$, the mean-field calculation  reveals  that the motion of the two spin states is separated and independent, with the motion of $I_\sigma$ for each component remaining completely localized  [see Fig.~\ref{fig9}(e)]. In the many-particle system, we observe the trajectories of $I_\sigma$ gradually approaching each other, with the rate of convergence accelerating as the number of particles $N$ decreases [Figs.~\ref{fig9}(a) and (c)]. However, for $\gamma$ not equal to zero, even at a very small value such as $\gamma=0.01$, the motion of $I_\sigma$ for the two spin-up and spin-down subsystems becomes clearly correlated in both the mean-field and full-quantum models, exhibiting the same oscillation amplitudes and consequently covering the same region of phase space, a key characteristic of measure synchronization\cite{96}. This investigation reveals that SOC greatly facilitates the synchronization dynamics between the spin-up and spin-down components. 
	
\begin{figure}[htp]
		\centering
		\includegraphics[width=8cm]{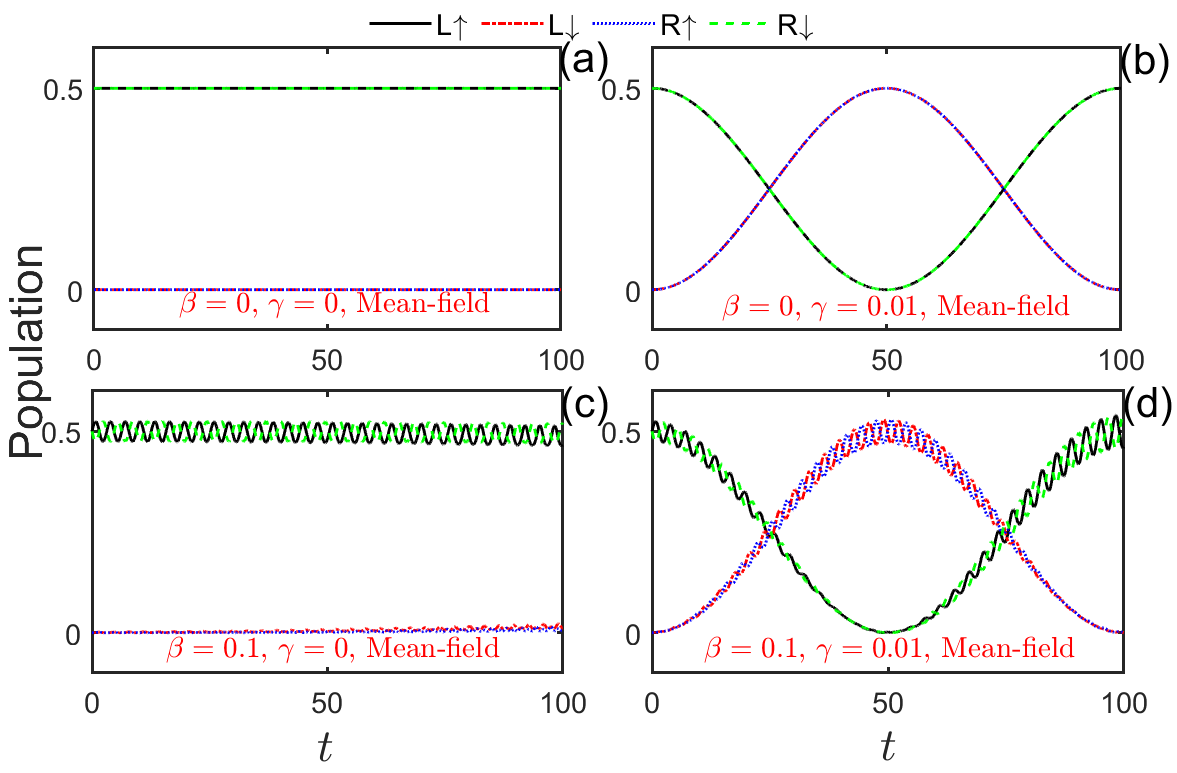}
		
		\caption{The time evolution of the populations on all four modes is shown for the mean-field model for \( \gamma = 0 \) (left column) and \( \gamma = 0.01 \) (right column), with $g=0.1$. Panels (a) and (b) correspond to the case with \(\beta = 0\), while panels (c) and (d) are for the case with \(\beta = 0.1\). The initial state is the same as that in Figs.~\ref{fig9} (e) and (f).}\label{fig10}
	\end{figure}

	For more detail, we present the time evolution of the populations for the corresponding four modes in the mean-field model with $\beta=0$ (first row) and $\beta=0.1$ (second row), as shown in Fig.~\ref{fig10}. By comparing the cases of $\beta=0$ and $\beta=0.1$, we clearly find that dissipation plays only a perturbative role in the dynamical evolution. Thus, the mean-field dynamics presented in Figs.~\ref{fig9}(e) and (f) can be concluded from the $\beta=0$ case. Furthermore, when $\beta=0$, it is found that the population distributions in the modes $\ket{L\uparrow}$ and $\ket{R\downarrow}$ behave identically throughout the evolution, and similarly, those in $\ket{L\downarrow}$ and $\ket{R\uparrow}$ are perfectly overlapped. The population dynamics in the absence of dissipation can be explained by the symmetry analysis of the mean-field model (\ref{29}) with $\beta=0$.
	
	We note that in the absence of dissipation, the mean-field model (\ref{29}) and its associated linear single-particle Hamiltonian ($g=0$) exhibit the following symmetry:
	\begin{equation}\label{con:4}
		\begin{split}\begin{aligned}
				\hat{S}=\begin{pmatrix}0&0&0&1\\0 & 0 & 1 &0\\0&1&0&0\\1&0&0&0\end{pmatrix}.
		\end{aligned}\end{split}
	\end{equation}
	Here, $\hat{S}$ represents the operator that flips the spin and interchanges the two wells. Starting from the initial state $|\psi(0)\rangle=\frac{1}{\sqrt{2}}(1, 0, 0, 1)^{\mathrm{T}}$, which obeys the $\hat{S}$ symmetry, it follows that the solution to the Schr\"odinger equation (\ref{28}) also  obeys the $\hat{S}$ symmetry, which yields $x_1(t) = x_4(t)$ and $x_2(t) = x_3(t)$ [accounting for  the overlapping of populations  in Figs.~\ref{fig10}(a) and (b)], and hence $\xi_{1}^{\prime} = \xi_{2}^{\prime} = \xi_{3}^{\prime} = \xi_{4}^{\prime}$. When $\gamma=0$, it is easy to verify that this initial state happens to be an eigenstate of the linear single-particle Hamiltonian. As the diagonal terms (i.e., the nonlinear terms) in the mean-field model (\ref{29}) are equal, that is, $\xi_{1}^{\prime} = \xi_{2}^{\prime} = \xi_{3}^{\prime} = \xi_{4}^{\prime}$, which can be gauged away by introducing an irrelevant global phase in the wavefunction, the $\hat{S}$ symmetry protects the motion pattern of the mean-field model, similar to that in the linear single-particle system, ensuring that modulus squared of the coefficients in the four modes remains unchanged during the evolution, as illustrated in Fig.~\ref{fig10}(a).
	\begin{figure}[htp]
		\centering
		\includegraphics[width=8cm]{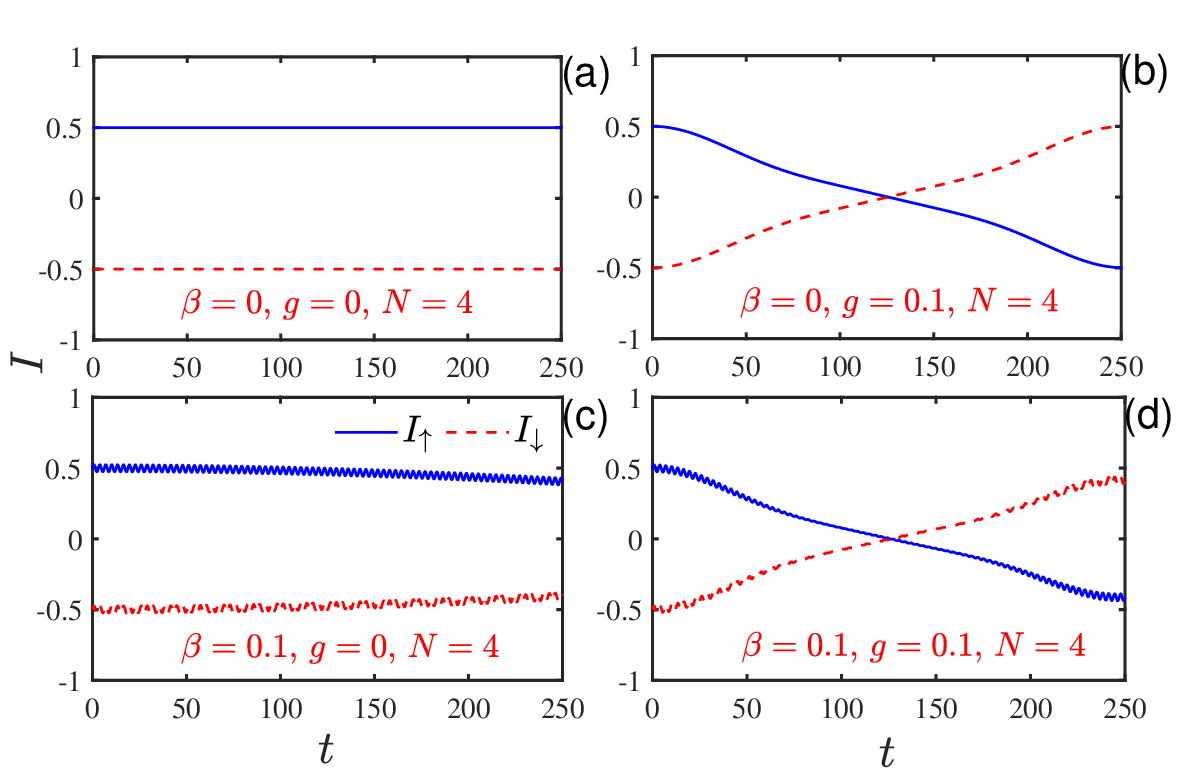}
		
		\caption{The time evolutions of \( I_\uparrow \) and \( I_\downarrow \) are analyzed for the full-quantum system with \( N = 4 \) and $\gamma=0$, for \( g = 0 \) (left column) and \(g = 0.1 \) (right column). (a) and (b)  are for \( \beta = 0 \), while (c) and (d)  are for \( \beta = 0.1 \). The initial coherent state is prepared as the same as in Fig.~\ref{fig9}.}\label{fig11}
	\end{figure}
	
	In Fig.~\ref{fig11}, we further investigate the effect of particle interaction on the synchronization dynamics in the full-quantum system with $N=4$ and $\gamma=0$. It is observed that the introduction of particle interactions causes the trajectories of the two spin states to change from being independent [Figs.~\ref{fig11}(a) and \ref{fig11}(c)] in single-particle [or equivalently, noninteracting many (few)-particle] quantum systems to intersecting and swapping with each other [Figs.~\ref{fig11}(b) and \ref{fig11}(d)] in interacting quantum systems. This is because the symmetrically protected eigenstates of the single-pariticle system  are destroyed by the particle interactions, leading to the emergence of measure synchronization. Comparison of the $\beta=0$ [Figs.~\ref{fig11}(a) and \ref{fig11}(b)] and $\beta=0.1$ [Figs.~\ref{fig11}(c) and \ref{fig11}(d)] cases also reveals that dissipation plays only a perturbative role in the dynamical evolution. From Figs.~\ref{fig9}, \ref{fig10}, and \ref{fig11}, it seems that when comparing the two different mechanisms—interactions and SOC—for causing measure synchronization to happen, the interactions are less effective in promoting synchronization than SOC. Due to the $\hat{S}$  symmetry, the synchronization dynamics are evidenced by the simultaneous yet out-of-phase oscillations of the spin-up and spin-down components, where $I_{\uparrow}=-I_{\downarrow}$.
	
	\section{CONCLUSIONS}
	\label{section 4}
	In summary, we have studied the quantum-classical correspondence in a non-Hermitian SO-coupled bosonic junction, where an effective decay term is introduced in one of the sites. It is analytically demonstrated that the mean-field dynamics of such a non-Hermitian SO-coupled bosonic junction can be described by a non-Hermitian discrete Gross-Pitaevskii equation with a classical Hamiltonian function. Our numerical studies examine the dependence of the symmetry-breaking threshold in the closely related $\mathcal{PT}$-symmetric system on the interaction strength, particle number, and SOC  strength. We find that when the effective SOC strength assumes a half-integer value, the symmetry-breaking threshold is zero regardless of the particle number and other system parameters. Through the study of atomic and spin population imbalances between two wells, we find that in the symmetry-breaking regime, the atoms eventually become self-trapped in the site without loss, whereas the spin population imbalance displays oscillatory behavior. The oscillatory behavior is associated with the appearance of multiple steady eigenstates, all sharing the same largest imaginary eigenvalues. We also find that the critical point of transition to atomic self-trapping from the Josephson oscillation depends on the particle number, and near the transition point, the breakdown of the mean-field and quantum correspondence is readily observed. While the quantum dynamics is less sensitive to the particle number when the SOC strength takes a half-integer value, atomic self-trapping occurs in all cases. Additionally, we reveal that both the atomic interaction and the SOC effects can cause synchronous periodic oscillations of the up and down spins, and the synchronous dynamics is protected by a type of symmetry associated with spin flipping and the interchange of the two wells.  Finally, we point out that the analytical procedure and framework developed in this work for deriving the classical Hamiltonian function and the associated discrete nonlinear Schrödinger equation are applicable to a broader class of extended non-Hermitian, $N$-particle, $M$-mode Bose-Hubbard systems. It is interesting to note that when extending to more modes or adding periodic driving, the mean-field dynamics may become chaotic, and studying the quantum-classical correspondence will be our future work.
	
	\acknowledgments
	The work was supported by the National Natural Science Foundation of China (Grants No. 12375022 and No. 11975110), the Natural Science Foundation of Zhejiang Province (Grant No. LY21A050002), Zhejiang Sci-Tech University Scientific Research Start-up Fund (Grant No. 20062318-Y), and  Jiangxi Provincial Natural Science Foundation (No. 20232BAB201008).
	
	\nocite{1}
	\bibliography{reference1}	
	
\end{document}